# $B_{12}C_2N$ Interstitial Boron subcarbonitride with peculiar magnetic properties: First principles investigations

Samir F Matar[*]

Lebanese German University, LGU, Sahel Alma, Jounieh, Lebanon.

[*]Formerly at the University of Bordeaux, CNRS, France

Email s.matar@lgu.edu.lb, abouliess@gmail.com

https://orcid.org/0000-0001-5419-358X

**Abstract**

*The subcarbides $B_{12}C_3$ and $B_{13}C_2$ known for their abrasive properties, can be structurally considered as carbon inserted rhombohedral $\alpha$-$B_{12}$ and expressed as $B_{12}\{C\text{-}C\text{-}C\}$ and $B_{12}\{C\text{-}B\text{-}C\}$. Using density functional theory DFT computations, the substitution of the central atom in the linear triatomic interstitials by N leads to an original new boron subcarbonitride $B_{12}C_2N$ ($B_{12}\{C\text{-}N\text{-}C\}$) identified as slightly more cohesive than the two subcarbides. While $B_{12}C_3$ is insulating with a $E_{Gap}\sim0.2$ eV, and $B_{13}C_2$ a weak metal with small density of state DOS at the Fermi level, $B_{12}C_2N$ ground state is a half-metallic ferromagnet with M= 1 $\mu_B$. From the equation of state, the magnetization is found unchanged over a broad volume range around equilibrium. Total and magnetic charge density projections, in accordance with charge transfer calculations, show the charge envelopes to be concentrated on {C:N:C} and the magnetic charge having the shape of a torus centered on N. The triatomic interstitials interact through terminal C's specifically with one of the two boron substructures of $\alpha$-$B_{12}$ forming "3B1…C-N-C…3B1"-like complex. Further syntheses and experimental characterizations are expected to extend the field of investigation of such an original class of materials.*

Keywords: Boron; DFT; p-elements; magnetism.

**Introduction**

Boron crystallizes in complex varieties among which rhombohedral α-boron, $rh.B_{12}$ with space group $R\bar{3}m$ N°166, is particularly interesting because of a large vacant space able to host up to three interstitials of the rhombohedron shown with the large void in Fig. 1a. A schematic of generic interstials X1 (1-fold at ½,½,½) and X2 (2-fold at $x,x,x$) along the diagonal is shown in Fig. 1b. Nevertheless a representation using hexagonal settings is much more illustrative with the arrangement of $B_{12}$ icosahedra shown in Fig. 1c and the tri-interstitial sites aligned along the $c$-hexagonal vertical axis. Also the X2 end-interstitials in the linear triatomic X2-X1-X2 interact with one of the two boron substructures, B1 nearest neighbor versus farther B2. The remarkable distances are shown in Fig. 1c for the present case study applied to linear C-N-C interstitials. The d(C-N)=1.30 Å, d(B1-C)=1.67 Å and the angle $\angle$ B1-C-B1 = 104.6° let suggest two flattened tetrahedra along $c$-direction with the expression "3B1-C-N-C-3B1" highlighted within a yellow circle.

Such interstital compounds are commonly observed with light elements such as boron, carbon, nitrogen, and oxygen, expressed with generic chemical formulae $B_{12}X2_2X1$. For instance $B_{12}C_3$ is well known for its abrasive properties [1] together with its corresponding boron rich $B_{13}C_2$ where extra B replaces the central C [2]. $B_{13}N_2$ is the only broron subnitride stoichiometry [3]. $B_6O$ or $B_{12}O_2X1$ has been recently considered for different light elements X1 stable new phases [4]. The works focused on the physico-chemical change of properties brought by such interstitials as the mechanical ones regarding enhanced resistance to compressibility as well as electronic and magnetic structure originality depending on the chemical nature of the interstitials [4,5]. Particularly, the unusual onset of magnetism on light $p$-elements is the result of subtle charge exchange within the complex entity shown above, as well as with the host matrix. Such studies were helped with calculations based on the well established quantum framework of the density functional theory (DFT) [6,7]. It is the purpose of this work to address such properties within weakly metallic $B_{13}C_2$ ($B_{12}C_2B$) where central boron is replaced by nitrogen, leading to a subcarbonitride $B_{12}C_2N$ that is identified as cohesive as $B_{13}C_2$ and $B_{12}C_3$ and characterized by the peculiar onset of a stable 1 $\mu_B$ magnetization in a half-metal ferromagnet ground state.

## 1. Computational framework

For the search of the ground structure of the newly proposed ternary subcarbonitride, geometry optimizations of the atomic positions and lattice parameters were carried starting from experimental data of $B_{12}C_2B$ [2]. Total energy calculations and electronic structure ware also carried out for the the two boron subcarbides binaries. For this purpose, we used the plane-wave VASP code [8,9]. Built within DFT, the code uses the projector augmented wave (PAW) method [9,10] for the atomic potentials. Accounting for DFT effects of exchange and correlation XC was operated within the generalized gradient approximation (GGA) [11]. The conjugate-gradient algorithm [12] was used in this computational scheme to relax the atoms onto energetic ground state. For both geometry relaxation and total energy calculations the tetrahedron method with Blöchl corrections [13] as well as a Methfessel-Paxton [14] scheme were applied. Brillouin-zone (BZ) integrals were approximated using a special k-point sampling of Monkhorst and Pack [15]. The optimization of the structural parameters was performed until the forces on the atoms were less than 0.02 eV/Å and all stress components less than 0.003 eV/Å$^3$. The calculations were converged at an energy cut-off of 500 eV for the plane-wave basis set concerning the **k**-point integration with a starting mesh of 6×6×6 up to 12×12×12 for best convergence and relaxation to zero strains.

For the purpose of establishing the equilibrium ground state properties, the equation of state EOS was derived from total energy calculations at different volumes around minima found from the geometry optimization. The resulting quadratic E(V) curve was fitted with 3$^{rd}$ order EOS [16]:

$$E(V) = E_0(V_0) + \frac{9}{8}V_0 B_0[(V_0/V)^{2/3} - 1]^2 + \frac{9}{16}B_0(B'-4)V_0[(V_0/V)^{2/3} - 1]^3$$

where $E_o$, $V_o$, $B_o$, and $B'$ are the equilibrium energy, the volume, the bulk modulus, and its pressure derivative.

The trends of charge transfers were assessed using a Bader charge analysis [17]. In this approach a partitioning is operated of the continuous electron density into regions bounded by the minima of the charge density. However such an analysis does not constitute a tool for evaluating absolute ionizations, but only trends. Bader's analysis is done using a fast algorithm operating on a charge density grid [18]. The results of computed charges are such that they lead to neutrality when the respective multiplicities are accounted for as shown in the tables.

Starting from the crystal parameters of the computed ground state structure, the site and spin projected density of states (DOS) were obtained using the full potential augmented spherical wave (ASW) method [20] and the GGA for the XC effects [11]. In the minimal ASW basis set, the outermost shells were chosen to represent the valence states and the matrix elements. They were constructed using partial waves up to $l_{max} + 1 = 2$ for B, C and N. Self-consistency was achieved when charge transfers and energy changes between two successive cycles were such as: $\Delta Q < 10^{-8}$ and $\Delta E < 10^{-6}$ eV, respectively. The BZ integrations were performed using the linear tetrahedron method within the irreducible rhombohedral wedge following Blöchl scheme [13].

2. Calculations and discussion of the results

   a- Energy trends and valence electron count

For the series of interstitial compounds based on $B_{12}$, a preliminary step was to examine them from the valence electron count (VEC) and interatomic charge transfers on one hand and to establish a comparative overview of the respective cohesive energies as averaged per atom, on the other hand. Focus was made on $B_{12}$ and the $B_{12}C_2X1$ (X1= B, C, N) chemically related series. The X1 = N member is proposed herein besides the experimentally known $B_{12}C_3$ and $B_{13}C_2$. Parameter-free, non-constrained total energies were obtained from successive self-consistent sets of calculations at an increasing number of k-points; then the cohesive energies were deducted from subtracting the energies of the atomic constituents. Table 1 presents the corresponding results. $B_{12}$ host characterized by VEC($B_{12}$)= $N_0$ = 36 corresponding to a closed shell, has a cohesive energy of -1.15 eV. It is also found as a small gap insulator. $B_{12}C_3$ adds up 12 more electrons and VEC($B_{12}C_3$)= 48 translating again a closed-shell electronic system and an insulator as shown hereacter in the development of the site projected electronic DOS.

$B_{12}C_3$ is expectedly more cohesive than $B_{12}$, due to establishing covalent B-C and C-C bonds (cf. generic Fig. 1b). The situation changes with $B_{13}C_2$ with VEC = $N_0$+11 = 47, i.e. one electron less than in $B_{12}C_3$ which then has a hole in the valence band VB, i.e. with the top of the VB band crossed by the Fermi level $E_F$ at low magnitude, leading a weak metal as shown in next sections (DOS). By replacing B at X1 by N, the sub-carbonitride $B_{12}C_2N$ has one more electron versus $B_{12}C_3$ leading to a paramagnet and eventually to a stable ferromagnetic ground state. It is found within range but slightly more cohesive than $B_{13}C_2$ (experimental compound).

The last column of Table 1 shows the Bader charges highlighting a negligible modification of B2 substructure versus B1. Each of the six B1 atoms loses less than one electron towards X2-X1-X2 which is negatively charged in all three interstitial compounds. For covalent $B_{12}C_3$ the lowest total charge exchange $\Sigma\delta\pm$ is observed with $\Sigma\delta\pm = 4.38$. $B_{13}C_2$ and $B_{12}C_2N$ have very close $\Sigma\delta\pm$ magnitudes with a slightly larger transfer from B1 to C-N-C ($\delta_{B1} =+0.94$ or $\Sigma\delta\pm=5.64$) explained by the bigger electronegativity of N versus B in C-B-C.

b- Ground state structure

Starting from $B_{13}C_2$ crystal data [2], B was replaced by N followed by full geometry relaxation of $B_{12}C_2N$ considering non spin polarized NSP configuration. The rhombohedral symmetry was found to be preserved throughout the calculation cycles. The structure results are given in Table 2 with the rhombohedral setting of $R\bar{3}m$ space group N°166; the corresponding hexagonal parameters are provided at the lower part of the table. Further calculations assuming spin polarization (SP) led to lowering of the energy from -105.73 eV (Table 1) to -106.93 eV ($\Delta E(SP-NSP) = -1.2$ eV) without change of the crystal parameters but a development of $1\mu_B$ (Bohr Magneton) magnetization was confirmed throughout the increasing precison of the BZ k-mesh. The interatomic distances are given for closest neighbors, highlighting the shortest distances within "3B1-C-N-C-3B1" shown circles in yellow in Fig. 1c.

c- Charge and magnetic densities and electron localization

In the post-treatment process of the ground state electronic structure, graphical analyses were applied to total charge density (CHGCAR) and to the magnetic charge density (CHGCAR_magn). In Fig. 2a the charge density is projected considering hexagonal settings, i.e. for $B_{36}C_6N_3$ stoichiometry for a clear representation especially for different projections of views. The charge density is found concentrated around C-N-C with yellow envelops exhibiting largely inflated central volume around most electronegative N and triangular-like shape for end C extending towards B1. More details are shown on the edge cuts representing each a quarter of the projection. The largest magnitude concentrates on N with inflated ellipsoidal shape and a smaller magnitude pointing to C-N bond, letting observe that N has non bonding charge together with charge shared with end two carbon atoms. The role played by the non-bonding charge is exhibited in Fig. 2b obtained from the projection of the magnetic charge density showing clearly that the magnetization is concentrated on nitrogen with a toroid shape, away from the C-N-C axis, i.e. carried by the non-bonded charge.

d- Energy-volume curves and volume change of magnetization

Further investigation of the peculiar aspect of magnetization onset is needed as regards changes with volume. Considering spin-polarized SP configuration, the energy-volume curve of $B_{12}C_2N$ is shown in Fig. 3a. With a quadratic shape, it can be fitted with the 3$^{rd}$ order Birch equation of states EOS expressed in the computational section. The fit values are given in the insert. The bulk modulus $B_0$ = 236 GPa (Giga Pascal) can be confronted with experimental value of 224 ± 15 GPa for α-boron [19]. The higher magnitude found for the subcarbonitride is the result of the occupation of the interstitial space and the bonding of the interstitials with boron discussed above. Regarding volume dependent magnetic properties, Fig. 3b shows the change with volume (pressure) of the magnetization. The calculated points are joined to serve as a guide for the eye. At very low volumes far from equilibrium ($V_{eq.}$ = 110.87 Å$^3$ ), magnetization vanishes and it rapidly jumps to M= 1 $\mu_B$ starting from a small volume of 80 Å$^3$, and remains stable at the saturation magnetization within a long range around equilibrium volume. In so far that decreasing volume translates increasing pressure, the observed results let expect a stable magnetic system with pressure.

e- Electronic density of states DOS

Using the experimental crystal data of $B_{12}C_3$ and $B_{13}C_2$ and the calculated ones for $B_{12}C_2N$, calculations were carried out to fully describe the role of each constituent in the electronic structure using the all electrons, full potential Augmented Spherical Wave (ASW) method [20].

Fig. 4a shows the site projected density of states DOS of $B_{12}C_3$. The energy along the *x*-axis is at $E_V$ designating the top of the valence band (VB) separated from the conduction band CB in a small insulating gap. Oppositely, the zero energy is at the Fermi level ($E_F$) crossing finite DOS in conducting $B_{13}C_2$ (Fig. 4b) and $B_{12}C_2N$ (Fig. 4c). These different electronic structure behaviors were already announced from the VEC count analyses.

In Fig. 4a, three main energy regions are identified: in the lowest energy part C1 and C2 together with B1 s-like states are found at -17 eV as well as at -4 eV where the corresponding $p_z$-states along the *c*-hexagonal axis are found. The VB from -10 up to $E_V$ is mainly dominated by B1 and B2 p-states which show similar skylines, pointing to their quantum mixing in forming the $B_{12}$ host matrix. Yet C1 and C2 p states are observed equally with lower magnitude due to the stoichiometry. The DOS features indicate a covalent lattice. Sharp empty C1, C2 and B1 states corresponding to the 3B1-C-C-C-3B1 complex mirror the lower VB part. In Fig. 4b, $B_{13}C_2$ shows similar quantum mixing features but the difference is in a continuous VB from -15 up to $E_F$, crossing the VB top at a low but significant DOS, whence the weakly metallic behavior. The less structured broad VB is due to a dominant B character and to the smaller global charge exchange of $\Sigma\delta\pm = 4.38$ versus $B_{12}C_3$ ($\Sigma\delta\pm = 5.52$) and $B_{12}C_2N$ ($\Sigma\delta\pm = 5.64$). Hence $B_{13}C_2$ can be considered as the most covalent within the series under consideration. Fig. 4c $B_{12}C_2N$ shows similar DOS features as $B_{12}C_3$ regarding the energy N(2s) followed by $p_z$-like mix of N, C and B1 at -15 eV. However a most interesting feature appears at the Fermi level crossed by a intense DOS peak composed mainly of p states of N with smaller intensity arising from C and B1. Such high-intensity DOS@$E_F$ signals instability of the electron system (mainly of N-2p) in such total spins configuration referred to within the Stoner theory of band ferromagnetism [21], whereby a stabilization of the electron system is expected upon accounting for two spin populations in spin-polarized SP calculations.

Indeed SP calculations considering two spin channels ↑ and ↓ resulted in a more stable state with $\Delta E(SP-NSP) = -1.34$ eV, a magnitude slightly larger than with geometry optimization procedure, but close enough to cast confidence on the results. The electronic SP band structure of $B_{12}CN_2$ along the main lines of the Brillouin zone is shown in Fig 5a. Green and blue lines correspond to spin ↑ while the blue lines are for spin ↓. The former, shifted to lower energy are called majority spins because of a larger electron population; the spin ↓ populations are called minority spins, the difference between the two provides the magnetic moment. One can see that the shift between the two spin populations is small over the lower part of the valence band VB, but the magnetic signature is mainly at the top corresponding mainly to nitrogen with a split of 0.2 eV observed between ↑ green bands at $E_F$, top of the VB and blue line bands of empty states in the CB. Detailed features are shown with the site and spin projected DOS at Fig. 5b. The DOS are now distributed in two subpanels for ↑ and ↓ spin populations. Similar features of DOS as in NSP calculations are shown for the p block in the energy window {-15 ; -1 eV}. Above this range, the major difference occurs with the splitting of p states as shown in the band structure panel. The large ↑ nitrogen DOS@$E_F$ and the ~0.2 eV energy gap observed in the ↓ spin DOS point to a half-metallic ferromagnet over a large volume range. While most of magnetization is carried by nitrogen as shown in the magnetic charge density in Fig. 2b, it needs to be mentioned that magnetic polarization is also carried to less extent by the atoms surrounding N especially within 3B1-C-N-C-3B highlighted in Fig. 1b; the whole summing up to M= 1 $\mu_B$. For the sake of completeness additional antiferromagnetic calculations with a doubling of the cell, one assigned SPIN UP magnetic subcell, the other SPIN DOWN magnetic subcell, led to zero magnetization but a much higher energy than SP configurations.

The sharp DOS shap indicating pronounced localization especially for the states lying on $E_F$ let suggest highly correlated states as in the transition metal oxide NiO, where regular DFT GGA XC functional becomes not sufficient for a correct account of the electronic system. In such cases the LDA+U / GGA+U calculations [22], where U designates the so-called Hubbard repulsive parameter, are useful for a correct account of the electronic structure. Complementary calculations using different magnituduses of U up to 7 eV led to a slight broadening of the DOS without change of the magnetization total value of 1 $\mu_B$ nor of the half- metallic ferromagnet property.

### 3  Conclusions

Based on starting experimental observations of α-dodecaboron subcarbide $B_{13}C_2$ expressed as an interstitial compound $B_{12}${C-B-C}, the substitution of central B by N has been shown to lead to an original new boron subcarbonitride identified as cohesive as the subcarbide and possessing the ground state property of a half-metallic ferromagnet. The magnetization of M= 1 $\mu_B$ is shown through the illustration of magnetic charge density taking the shape of a torus centered on N in {C-N-C} and arising mainly from non-bonding N electrons. We note however that DFT calculations are implicit of 0K, not accounting for temperature which has the effect of potentially bringing disorder in the long range magnetic order with the likely transformation to a paramagnet. Further planned synthesis and experimental characterizations are expected to extend the field of investigation of such an original class of materials. As in the other two binary subcarbides $B_{12}${C-C-C}, $B_{12}${C-B-C}, the chemical system is stabilized thanks to negatively charged triatomic linear interstitial, here {C-N-C} interacting with one of the two boron substructures of α-$B_{12}$ forming "3B1…C-N-C…3B1"-like complex illustrated from charge density. From energy calculations of different configurations, the ground state is identified as ferromagnetic with half-metallic property.

TABLES

Table 1. Cohesive energies of rhombohedral $B_{12}$ -based compounds,

| System | $E_{Tot.}$ | $E_{coh.}$/at. | VEC* | Bader, $\delta_{B2} \sim 0$ |
|---|---|---|---|---|
| $B_{12}$ | -80.49 | -1.15 | 36 (NM) | 0 (covalent networks) |
| $B_{13}C_2$ | -106.35 | -1.35 | 36+11=47 | $\delta_B$:+3; $\delta_{C2}$: -4.26; $\delta_{B1}$: +0.92. $\Sigma\delta\pm$=5.52 |
| $B_{12}C_3$ | -108.80 | -1.48 | 36+12=48 | $\delta_{C1}$:-2.48; $\delta_{C2}$: -0.93; $\delta_{B1}$: +0.73. $\Sigma\delta\pm$=4.38 |
| $B_{12}C_2N$ | -106.93** | -1.36 | 36+ 13=49 | $\delta_N$:-3.15; $\delta_{C2}$: -1.23; $\delta_{B1}$: +0.94. $\Sigma\delta\pm$=5.64 |

Energies of the atoms (eV): E(B)=-5.56. E(C ) =-6.48;  E(N)= -6.8

*VEC: Valence Electron Count.  **Spin polarized ground state energy

Table 2. $B_{12}C_2N$. Calculated crystal data starting from experimental ($B_{12}C_2B$) between brackets

$a_{rh}$ = 5.192 Å (5.198 Å) Å; α =65.42°(65.62)°

| Atom | Wyckoff | x | y | z |
|---|---|---|---|---|
| $B_1$ | *6h* | 0.808(0.804) | x | 0.309 (0.315) |
| $B_2$ | *6h* | 0.006 (0.005) | x | 0.330 (0.329) |
| N (B) | *1b* | ½ | ½ | ½ |
| C | *2c* | 0.393 (0.385) | x | x |

Calculated Shortest distances in $B_{12}C_2N$
d(C-N)=1.31 Å
d(C-B1) =1.67 Å
d(B1-B1)= 1.74 Å
d(B2-B2) =1.82 Å.
d(B1-B2) =1.79 Å.

N.B. In hexagonal setup: $a_{hex.}$=5.611 Å, $c_{hex.}$=12.172 Å.

N (0,0, ½); C (0,0, 0.393); B1 (0.166, 0.332, 0.64167); B2 (-0.108, -0.216, 0.114)

FIGURES

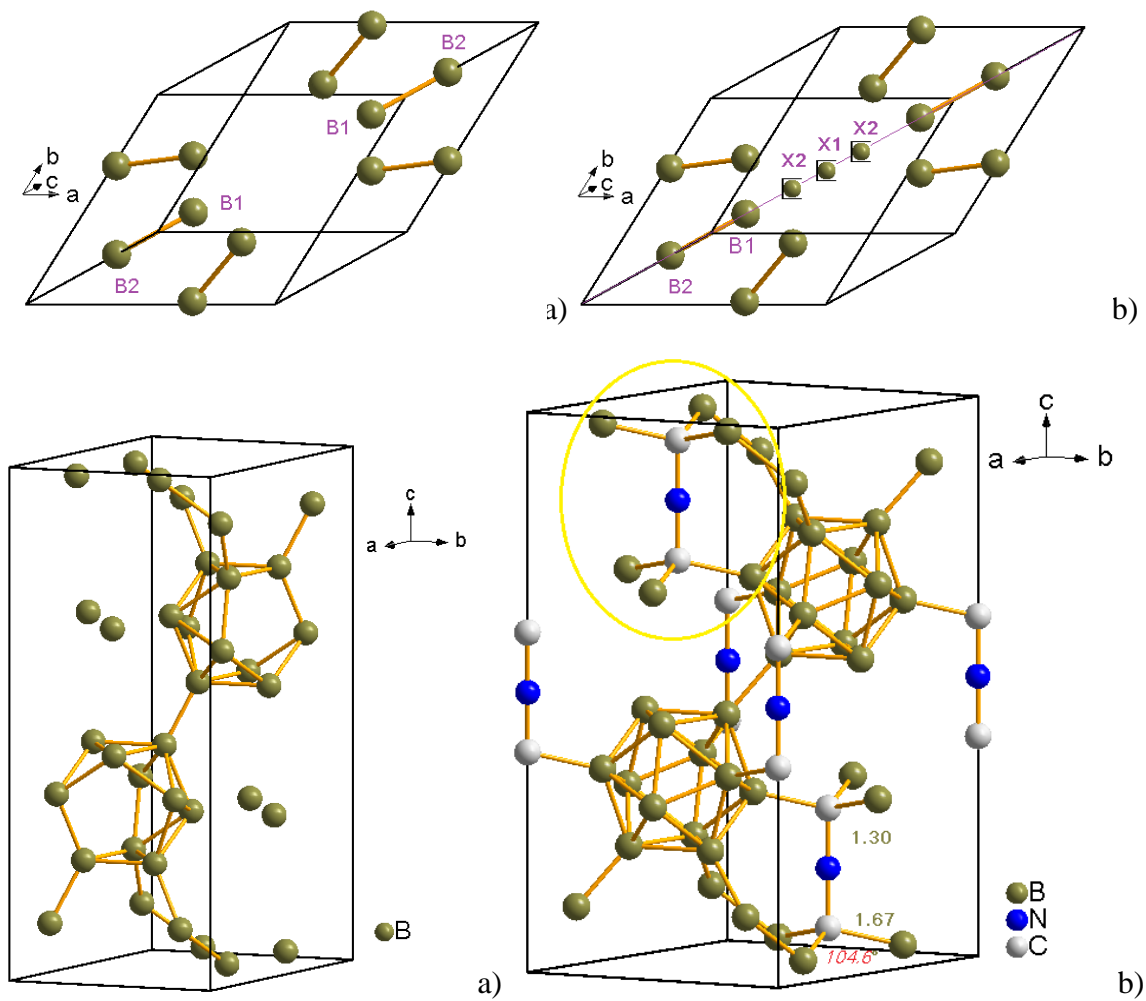

Figure 1. a) *rh*.B$_{12}$C showing the central empty space, b) interstials occupying two different crystal sites 1-equivalent X1 (½ ½ ½) and 2-equivalent X2 (x,x,x). c) *rh*.B$_{12}$, and d) *rh*.B$_{12}$C$_2$N represented with hexagonal axes; hee X1 is occupied by N and X2 by C.

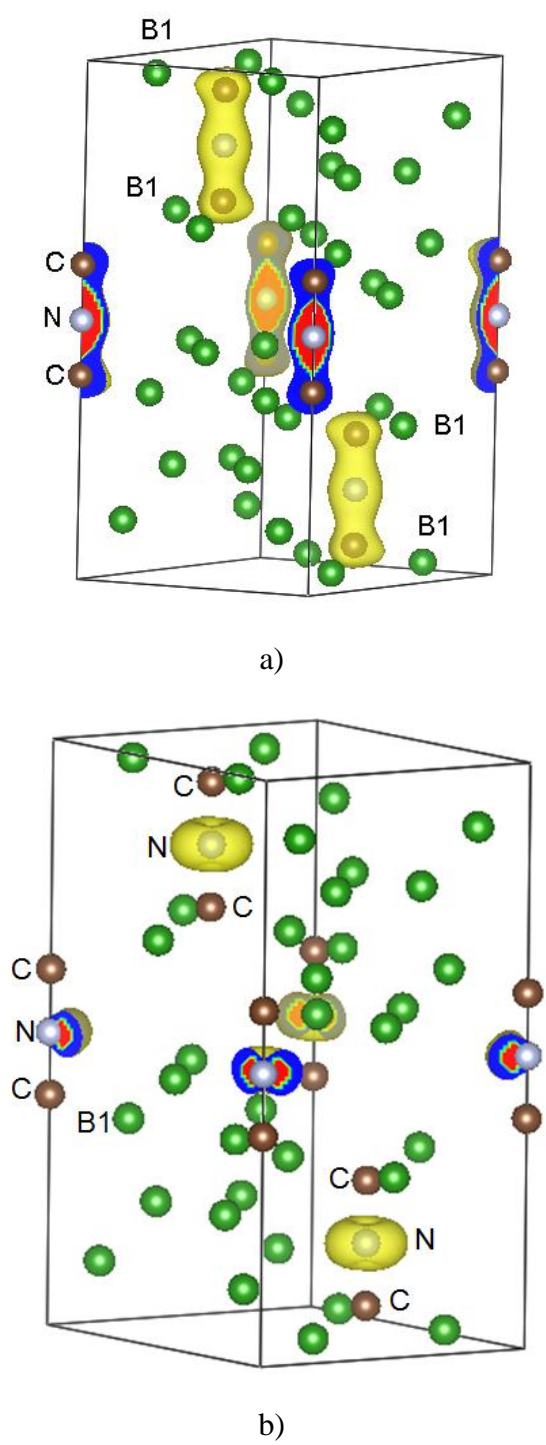

Figure 2. Charge densities in *rh*.B$_{12}$C$_2$N (hexagonal axes): a) Total charge density; b) Magnetic charge density exhibited around N with 1 $\mu_B$ magnetic moment.

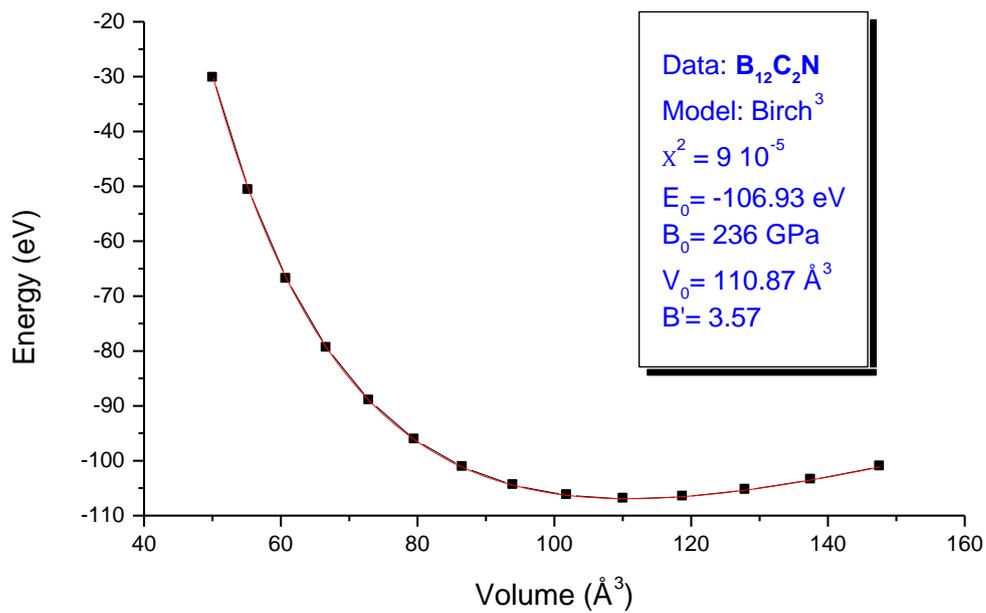

a)

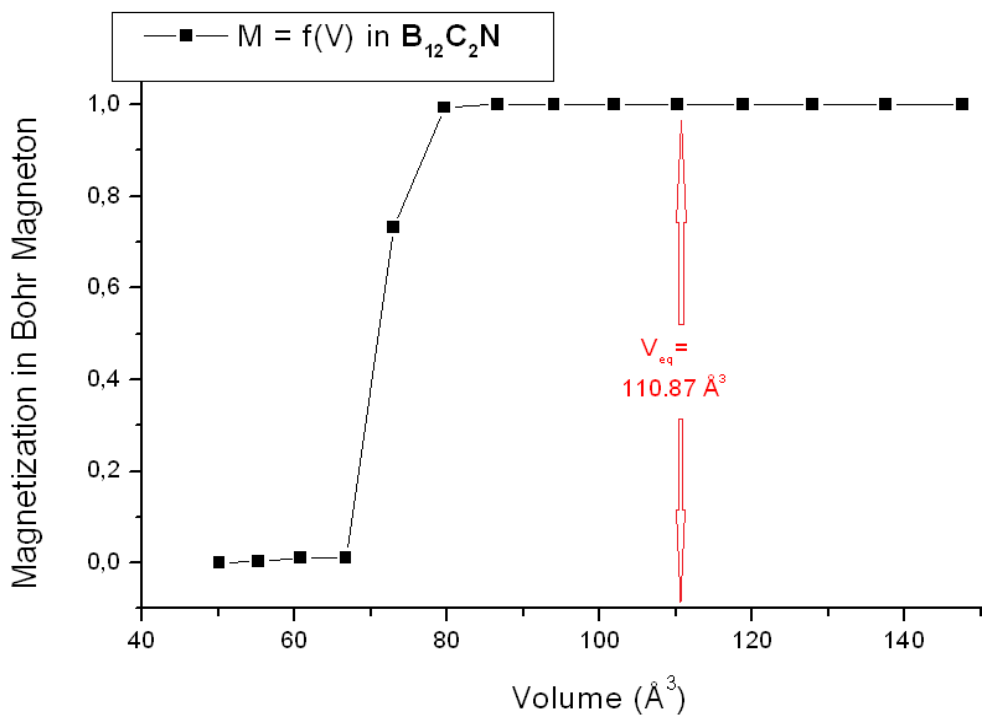

b)

Figure 3. $B_{12}C_2N$: a) Energy-volume curve and fit values in the insert; b) Volume change of the magnetization found to be stable at 1 $\mu_B$ over a broad range of volume around equilibrium.

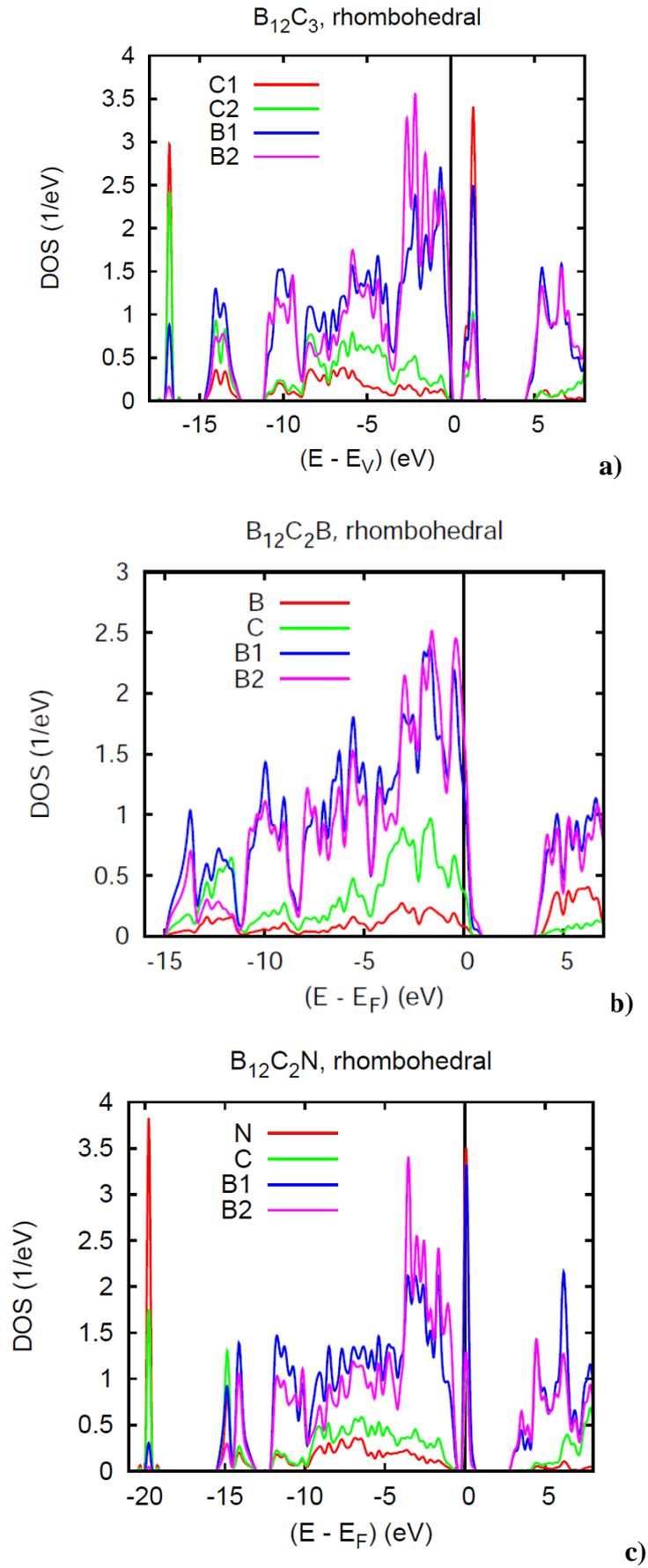

Figure 4. Non magnetic calculations of the electronic density of states (DOS).

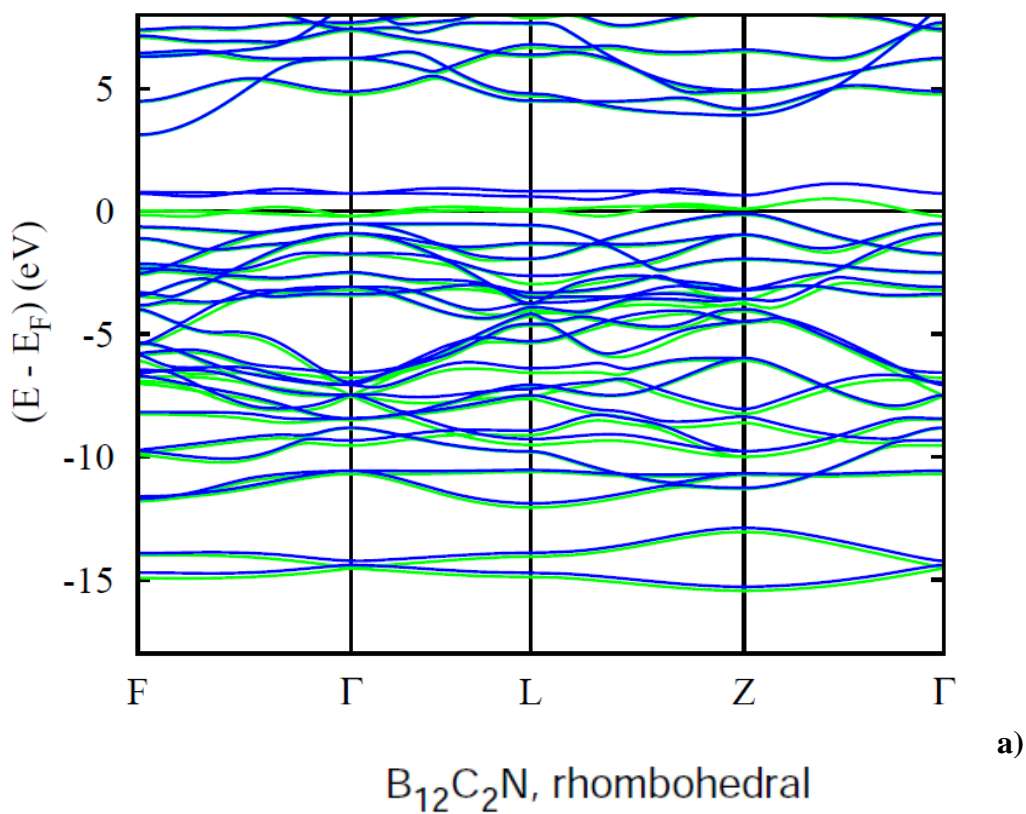

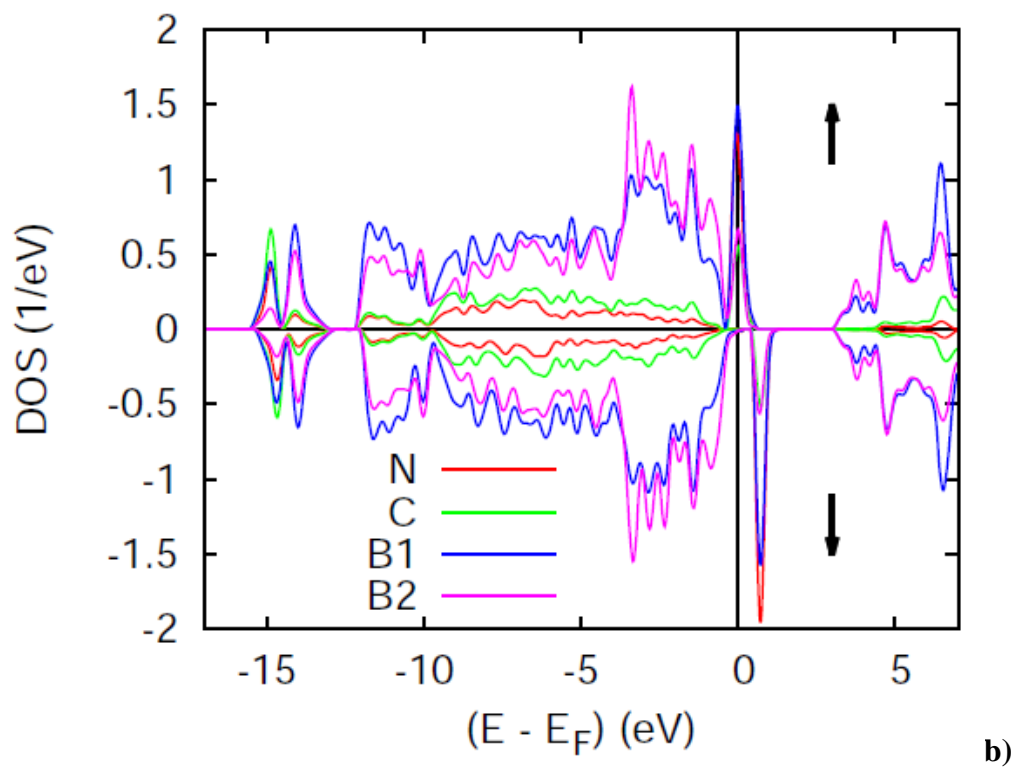

Figure 5. $B_{12}C_2N$ spin polarized magnetic SP calculations. a). Electronic band structures with green bands for majority spins ↑, and blue lines for minority spins ↓. b) Site and spin projected DOS.

17